\documentclass[twocolumn,amsmath,amssymb,superscriptaddress,aps,prl]{revtex4-1}

\usepackage{color}
\usepackage{graphicx}
\usepackage{amssymb}
\usepackage{amsmath}
\usepackage{mathrsfs}
\usepackage[normalem]{ulem}
\newcommand{\ud}{\mathrm{d}}

\newcommand{\pdern}[3]{\frac{\partial^#3#1}{\partial#2^#3}}
\newcommand{\pder}[2]{\frac{\partial#1}{\partial#2}}

\let\Eps\varepsilon
\newcommand{\aasays}[1]{\textcolor{black}{#1}}

\usepackage[breaklinks=true]{hyperref}
\graphicspath{ {./figures/} {./}}

\begin{document}

\title{Stabilization of unsteady nonlinear waves by phase space manipulation
}

\author{Alexis Gomel}
\affiliation{GAP, Universit{\'e} de Gen{\`e}ve, Chemin de Pinchat 22, 1227 Carouge, Switzerland}
\affiliation{Institute for Environmental Sciences, Universit{\'e} de Gen{\`e}ve, Boulevard Carl-Vogt 66, 1205 Gen{\`e}ve, Switzerland}

\author{Amin Chabchoub}
\affiliation{Centre for Wind, Waves and Water, School of Civil Engineering, The University of Sydney, NSW 2006, Australia}
\affiliation{Disaster Prevention Research Institute, Kyoto University, Kyoto 611-0011, Japan}

\author{{Maura Brunetti}}
\affiliation{GAP, Universit{\'e} de Gen{\`e}ve, Chemin de Pinchat 22, 1227 Carouge, Switzerland}
\affiliation{Institute for Environmental Sciences, Universit{\'e} de Gen{\`e}ve, Boulevard Carl-Vogt 66, 1205 Gen{\`e}ve, Switzerland}

\author{Stefano Trillo}
\affiliation{Department of Engineering, University of Ferrara, via Saragat 1, 44122, Ferrara, Italy}

\author{J{\'e}r{\^o}me Kasparian}
\email{jerome.kasparian@unige.ch}
\affiliation{GAP, Universit{\'e} de Gen{\`e}ve, Chemin de Pinchat 22, 1227 Carouge, Switzerland}
\affiliation{Institute for Environmental Sciences, Universit{\'e} de Gen{\`e}ve, Boulevard Carl-Vogt 66, 1205 Gen{\`e}ve, Switzerland}
\date{\today}

\author{Andrea Armaroli}
\email{Present address: PhLAM - Laboratoire de Physique des Lasers, Atomes et Mol\'ecules, IRCICA, 50 avenue Halley, 59658 Villeneuve d'Ascq, France. andrea.armaroli@univ-lille.fr}
\affiliation{GAP, Universit{\'e} de Gen{\`e}ve, Chemin de Pinchat 22, 1227 Carouge, Switzerland}
\affiliation{Institute for Environmental Sciences, Universit{\'e} de Gen{\`e}ve, Boulevard Carl-Vogt 66, 1205 Gen{\`e}ve, Switzerland}

\begin{abstract}
We introduce a dynamic stabilization scheme universally applicable to unidirectional nonlinear coherent waves.
By abruptly changing the waveguiding properties, the breathing of wave packets subject to modulation instability can be stabilized as a result of 
the abrupt expansion a homoclinic orbit and its fall into an elliptic fixed point (center).
\aasays{We apply this concept to the nonlinear Schr{\"o}dinger equation framework and show that an Akhmediev breather envelope, which is at the core of Fermi-Pasta-Ulam-Tsingou recurrence and extreme wave events}, can be \textit{frozen} into a steady periodic (dnoidal) wave by a suitable variation of a single external physical parameter. We experimentally demonstrate this general approach in the particular case of surface gravity water waves propagating in a wave flume with an abrupt bathymetry change. Our results highlight the influence of topography and waveguide properties on the lifetime of nonlinear waves and confirm the possibility to control them.
\end{abstract}

\maketitle

The parametric stabilization of unstable dynamics is a fascinating and long-standing problem, the paradigmatic example being the Kapitza pendulum \cite{Kapitza1951}, \textit{i.e.}, the dynamic stabilization of a pendulum around its inverted position by fast oscillating its pivot. 

Dynamic stabilization is still effective for nonlinear and dispersive waves which are intrinsically infinite-dimensional, unlike nonlinear control theory and feedback schemes. Applications range from dispersion management in fiber laser and communications \cite{Turitsyn2014} to control of nonlinear waves in many-body quantum physics \cite{Hoang2013}, diffractive optics \cite{Rizza2013}, matter waves \cite{Oberthaler2003}, or water waves \cite{Arie2017}. However, dynamic stabilization requires a spatially {\em extended} periodicity, and alternative stabilization and control schemes of nonlinear waves are needed \cite{Marcucci2019}.

Here, we introduce theoretically and validate experimentally such a nonlinear wave stabilization based on abruptly changing the propagation conditions, expanding a phase-space trajectory homoclinic to a saddle point \cite{ArnoldMMCM,Strogatz2015}. 
Generically, this trajectory contains a family of closed orbits, converging to a single point known as center. The phase-space manipulation stabilizes the system evolution around the center, suddenly \emph{freezing} the growth stage of a breather wave envelope at its peak height.

Unlike Kapitza or feedback schemes, such an expansion is induced by a controlled, {\em local and abrupt} variation of a single parameter affecting both the nonlinearity and the dispersion of the wave system. As an example, we apply this concept to unidirectional water surface gravity waves subject to the ubiquitous phenomenon of modulational instability (MI) of Stokes waves or Benjamin-Feir instability \cite{Benjamin1967,Zakharov2009,Dudley2019,Waseda2020}.  The evolution of such unstable waves can be described 
by the universal nonlinear Schr{\"o}dinger equation (NLSE) \cite{Zakharov1968}. MI entails the exponential growth of a slow modulation on top of a carrier wave of uniform amplitude possibly yielding to the formation of extreme waves. Remarkably, the continuation of MI in the fully nonlinear (strongly depleted) stage as modeled by Akhmediev breathers (ABs) is equivalent to  a homoclinic pendulum-like phase-space structure \cite{akhmediev1985generation,Ablowitz1990,Moon1990,Trillo1991c,Mussot2018}, where the background behaves as a saddle point, while two centers are represented by two out of phase stationary periodic wave envelopes, the dnoidal solutions of the NLSE \cite{Yuen1982,Magnani2020}. The unstable AB orbit  describes the amplification of sidebands up to a peak and the asymptotic return to the background \cite{Akhmediev1986}. Since it separates two qualitatively different types of periodic evolutions undergoing Fermi-Pasta-Ulam-Tsingou (FPUT) recurrences, the AB is a separatrix in the wave system phase-space~\cite{Akhmediev2001e,Kimmoun2016,Mussot2018}.

We demonstrate the possibility to stabilize such an unstable homoclinic orbit by matching it to one of the steady dnoidal solutions. The matching is strictly forbidden by the Hamiltonian structure of the NLSE for unperturbed MI evolutions. \aasays{Instead, we parametrically perturb the system by abruptly (\textit{i.e.}, faster than the MI characteristic distance)} increasing the water depth and, thus, changing the dispersion and nonlinearity experienced by the envelope. This causes a strong dilation of the AB orbit at its apex and, ideally, the fall of the trajectory over the center (dnoidal envelope). 
This blocks the FPUT recurrence and \emph{freezes} the breather at its peak.

The proposed \emph{separatrix dilation} is somehow opposite to the common phenomena of wave shoaling responsible for the increase of wave amplitude, typical for the depth decrease in coastal areas \cite{Zabusky1971,Djordjevic1978,Dutykh2011,Trulsen2012}. On the other hand, an increase of water depth in the direction of wave propagation can still occur in the ocean, mostly in surf zones like sandbars and coral reefs. The present mechanism can also occur where the NLSE provides a leading-order description of nonlinear MI, such as Bose-Einstein condensation \cite{Everitt2017} and optics, where quasi-stabilization has been interpreted in terms of solitons  \cite{Bendahmane2014a}. 
Indeed, this approach can be extended  to other models with a homoclinic structure \cite{Ercolani1990}, and even to settings such as parametric resonance described by strongly non-integrable models \cite{Conforti2016}.

A NLSE-like equation was derived for the one-dimensional and uni-directional evolution of the envelope of surface water waves on an uneven bottom of depth $h$ at frequency $\omega=\sqrt{gk\sigma}$, with $\sigma\equiv\tanh{\kappa}$ and $\kappa\equiv k h$, $k$ being the local wavenumber, which varies with $h$, while $\omega$ is fixed~\cite{Djordjevic1978}.
The slope of the depth step in the propagation direction $x$ should be sufficiently small to prevent wave reflections due to wavenumber mismatches: $h'(x)=\mathcal{O}( \Eps^2)$, with $\Eps\equiv k a$ the wave steepness, $a$ being the carrier wave amplitude. 	
\aasays{
Applying the method of multiple scales up to $\mathcal O( \Eps^3)$ to the inviscid irrotational water wave problem yields the evolution equation \cite{Djordjevic1978,Mei2005} 
\begin{equation}
i\pder{V}{\xi} + \beta\pdern{V}{\tau}{2} - \tilde\gamma|V|^2V = 0,
\label{eq:DReq2}
\end{equation}
where $V(\xi,\tau)$ is the shoaling-corrected envelope of the free surface elevation \cite{Onorato2011,Armaroli2020}, $\xi\equiv\Eps^2 x$, and $\tau\equiv \Eps\left[\int_0^x \frac{\ud \zeta}{c_\mathrm{g}(\zeta)}-t\right]$ ($t$ being the physical time) are the coordinates in a frame moving at the envelope group velocity, $c_\mathrm{g} \equiv \pder{\omega}{k}= \frac{g}{2\omega}\left[\sigma + \kappa(1-\sigma^2)\right]$.}  

Here $\tilde\gamma \equiv \gamma \frac{c_\mathrm{g}(\xi=0)}{c_\mathrm{g}(\xi)}$
is the shoaling-induced correction of the standard nonlinear coefficient $\gamma$, and $\beta$ is 
the group-velocity dispersion. They only depend on $\kappa$ \cite{Mei2005}, with $\beta<0$ regardless of $\kappa$ (only surface gravity waves are considered \cite{Mei2005}) and $\tilde\gamma\ge0$ for $\kappa\ge 1.363$, so that $\beta\tilde\gamma<0$
(in this focusing regime $c_\mathrm{g}$ monotonically decreases, and shoaling only increases slightly the effective nonlinearity $\tilde\gamma$, see Supplemental Material S1 \cite{SM2020}).

The NLSE~\eqref{eq:DReq2} conserves only the mass $N\equiv\int_{-\infty}^{\infty}{|V|^2\ud \tau }$ and the momentum $P \equiv \mathrm{Im}\left\{\int_{-\infty}^{\infty}{ V^*\pder{V}{\tau}}\ud \tau\right\}$, which we use in our numerical simulations to ensure the integration precision. Moreover, we introduce the quantities $A\equiv V/V_0$, $X \equiv \xi/L_\mathrm{nl}$, $T \equiv \tau/T_\mathrm{nl}$, \aasays{where $V_0$ is the amplitude of the input plane wave (carrier), while $L_\mathrm{nl}=1/(\tilde\gamma V_0^2)$ and $T_\mathrm{nl}=\sqrt{2 |\beta| L_\mathrm{nl}} = \sqrt{\frac{2|\beta|}{\tilde\gamma V_0^2}}$ are the associated characteristic nonlinear length and temporal scales, respectively}. This allows us to cast Eq.~\eqref{eq:DReq2} into the dimensionless focusing NLSE
\begin{equation}
i\pder{A}{X} - \frac{1}{2}\pdern{A}{T}{2} - |A|^2A = 0.
\label{eq:NLS}
\end{equation}

We let the depth increase from $h^0$ to $h^\infty>h^0$ over a a distance $L_\mathrm{step}$, with $\frac{h^\infty-h^0}{\Eps^2}\ll L_\textrm{step} \ll L_\mathrm{nl}$, to prevent spurious reflections \cite{Djordjevic1978}, while remaining essentially local compared to the envelope scale of variation $L_\mathrm{nl}$.
The normalization of Eq.~\eqref{eq:NLS} changes from before to after the bathymetry change. Assuming a fixed mass $N$ (shoaling being negligible), two different families of solutions of Eq.~\eqref{eq:NLS} can be matched across the change
(henceforth, superscripts $0$ and $\infty$ denote the physical quantities before and after the change).

\begin{figure}[ht]
\centering
\includegraphics[width=0.45\textwidth]{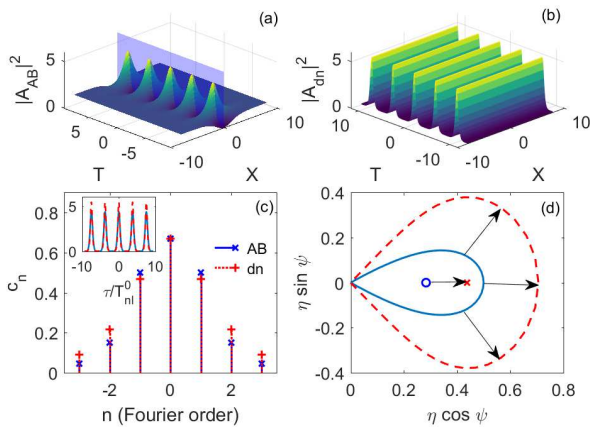}
\caption{Principle of AB conversion into a dnoidal solution of the NLSE. (a) Space-time-evolution of the AB for a normalized detuning $\Omega^0=1.67$; (b) dnoidal solution for normalized detuning $\Omega^\infty=1.34$; 
(c) Best matching of Fourier coefficients of the AB [at the peak distance, blue shading in panel (a)] to the dnoidal solution. Inset: superimposed time profiles;
(d) Phase plane trajectories of AB and dn-oidal for $\Omega^0$ (blue solid line and circle) and $\Omega^\infty$ (red dashed line and cross). 
Arrows: effect of separatrix dilation.
}
\label{fig:principle}
\end{figure}

First, we consider the AB solution \cite{Akhmediev1986}
\begin{equation}
	A_\mathrm{AB}(T,X) = \left[1+\frac{\frac{(\Omega^0)^2}{2}\cosh b X +i b \sinh{b X}}{\sqrt{1-\frac{(\Omega^0)^2}{4}}\cos \Omega^0 T-\cosh b X}\right]e^{i X},
\label{eq:AB}
\end{equation}
where $\Omega^0$ is the initial normalized MI sideband detuning and $b \equiv \Omega^0\sqrt{1-\frac{(\Omega^0)^2}{4}}$ the linear MI gain. This solution exists only for $0\le\Omega^0\le 2$, is periodic in $T$ and evolves in $X$ connecting two homogeneous plane wave states of unit amplitude at $X\to\pm\infty$ [Fig.~\ref{fig:principle}(a)]. It thus corresponds to the separatrix of infinite-dimensional NLSE.

Second, recall the dnoidal solutions \cite{Yuen1982},
\begin{equation}
A_\mathrm{dn}(T,X;m) = \alpha \,\mathrm{dn}\left[\alpha T;m \right]e^{i \chi^2 X},
\label{eq:dnoidal}
\end{equation}
where $\alpha = \chi\sqrt{\frac{2}{2-m^2}}$, and $\chi$ a constant to be determined. The parameter $0<m<1$ implicitly defines the solution period: $T_\mathrm{dn} = \frac{2 K}{\alpha}$, where $K\equiv K(m)$ is the complete elliptic integral of the first kind \cite{Byrd1971}. This solution has a steady amplitude profile  and generalizes the soliton solution for $T$-periodic boundary conditions [Fig.~\ref{fig:principle}(b)]. It is the infinite-dimensional counterpart of a center in a Hamiltonian system. 

We seek $m$ that matches the breather solution $A_\mathrm{AB}(T,X)$ to a steady profile $A_\mathrm{dn}(T,X;m)$ at a given stage $X$ of the evolution. This will stabilize ("\textit{freeze}") a strongly modulated nonlinear state. 

Considering the phase invariance of the NLSE and the realness of the AB at its peak position $X=0$, we choose  $A^0_\mathrm{AB}\equiv -A_\mathrm{AB}(T,0)$ to have positive maxima and negative minima ($A_\mathrm{AB;max,min}^0 = 1\pm \sqrt{4-(\Omega^0)^2}$), inset Fig.~\ref{fig:principle}(c) corresponding to the shaded blue plane of Fig.~\ref{fig:principle}(a). 
 We expand this real-valued wave form in Fourier series $A^0_\mathrm{AB}(T) =  c_0^0 + \sum_{n\neq 0} c_n^0 e^{ i n\Omega^0 t}$, \cite{Bendahmane2015} with 
\begin{equation}
	 c_0^0=(\Omega^0-1);\: c_n^0 = \Omega^0  \left(\frac{2-\Omega^0}{2+\Omega^0}\right)^\frac{|n|}{2},
\label{eq:ABFourier}
\end{equation}
[Fig.~\ref{fig:principle}(c)].
The  dnoidal profile that best matches $A^0_\mathrm{AB}$ must first be real-valued like the AB. Therefore we take $A^\infty_\mathrm{dn}(T)\equiv A_\mathrm{dn}(T,0)$. Second, the maxima $\alpha\ge 0$ at $T= k T_\mathrm{dn}$ and minima $\alpha\sqrt{1-m}\ge 0$ at $T = T_\mathrm{dn}/2 + k T_\mathrm{dn}$ of  $A_\mathrm{dn}(T,X)$ must coincide to those of the AB. Third, neglecting shoaling, 
the conservation of $N$ implies $\alpha^2=K/E$ \cite{Byrd1971}, where $E\equiv E(m)$ is the complete elliptic integral of the second kind. Finally, the normalized detuning in the MI band corresponding to a particular dnoidal solution is derived by well-known formulas \cite{Byrd1971}, as 
$
\Omega^\infty \equiv \frac{\pi\alpha}{K} = \pi [K E]^{-\frac{1}{2}}.
$

The problem is thus reduced to finding the value of $\Omega^\infty$  that best matches $A^0_\mathrm{AB}(T)$  to $A^\infty_\mathrm{dn}(T)$. The latter expands in Fourier series:
\begin{equation}
c_0^\infty  = \frac{\Omega^\infty}{2};\; c_n^\infty= \Omega^\infty \  \frac{q^{|n|}}{1+q^{2|n|}},
\label{eq:dnFourier}
\end{equation}
with $q\equiv q(m)$ the elliptic \textit{nome}.

\aasays{We require that the main (continuous) components are equal, \textit{i.e.}, $c_0^0 = c_0^\infty$. The comparison of Eqs.~\eqref{eq:ABFourier} and \eqref{eq:dnFourier} yields}
\begin{equation}
\Omega^\infty = 2(\Omega^0-1),
\label{eq:freqmatch}
\end{equation}
providing the main theoretical result of our work. It simply links the two normalized pulsations across the depth change for optimally matching an AB to a steady dnoidal envelope.  Clearly, the envelope matching requires that the physical sideband detuning $f_m$ remains the same, whereas in Eq.~\eqref{eq:freqmatch} the pulsations $\Omega^{0,\infty}=2\pi f_m T_\mathrm{nl}^{0,\infty}$ differ on the two sides ($0,\infty$) because of the change in $T_\mathrm{nl}^{0,\infty}$, which accounts for the local depth. Thus, Eq.~\eqref{eq:freqmatch} is equivalent to $T_\mathrm{nl}^\infty=2 T_\mathrm{nl}^0-(\pi f_\mathrm{m})^{-1}$ allowing to determine $\kappa^\infty$ given $\kappa^0$. \aasays{Note also that $\Omega^\infty\ge\Omega^0$ for $0\le\Omega^0\le 2$: this is consistent with the requirement  $h^\infty>h^0$, because $T_\mathrm{nl}$ decreases monotonically with $\kappa$ and $h$, see the Supplemental Material S1.
 }

Figure~\ref{fig:principle}(c) compares the spectra of the AB and the dnoidal, when Eq.~\eqref{eq:freqmatch} is fullfilled. The sidebands ($n \ge 1$) match satisfactorily. Small unavoidable discrepancies induce small oscillations around the dnoidal (matching more than one $c_n$ is possible for the trivial case  $\Omega^0=\Omega^\infty=2$ only, \textit{i.e.}, vanishing jump and MI band-edge).
This is also obvious from the phase space representation of the matching process [(Fig.~\ref{fig:principle}(d)], where the variables $(\psi,\eta)$ are, respectively, the relative phase and sideband fraction of the two families of solutions (See \cite{Trillo1991c,Conforti2016,Vanderhaegen20} and the Supplemental Material S2 \cite{SM2020}).
The optimal jump [Eq.~\eqref{eq:freqmatch}] leads the separatrix apex ($\psi=0$) before the jump (blue solid line) to closely approach the center (red cross) standing for the dnoidal after the phase space dilation induced by the jump. Indeed the non-perfect superposition of the (blue) separatrix apex and the (red) center is responsible for small oscillations around the dnoidal after the jump. \aasays{These small oscillations around the maximum breather compression point are still nonlinear, because the energy is periodically exchanged between different sideband pairs.} Note that this approach can be adapted to near-separatrix conditions, as detailed in Supplemental Material S3~\cite{SM2020}. 

Our approach establishes that AB freezing is favored for $\sqrt{3}<\Omega^0<2$, since $A_\mathrm{AB}^0$ stays positive like $A_\mathrm{dn}^\infty$
(for $0\le\Omega^0\le \sqrt{3} $ the AB takes negative values, inaccessible to the dnoidal family). 
However, in Fig.~\ref{fig:principle} and in the experiment, we operate slightly below $\Omega^0=\sqrt{3}$ to increase the MI gain, but still the temporal profiles show a very good matching
[inset of Fig.~\ref{fig:principle}(c)].

%

\begin{figure}[!ht]
\centering
\includegraphics[width=\columnwidth]{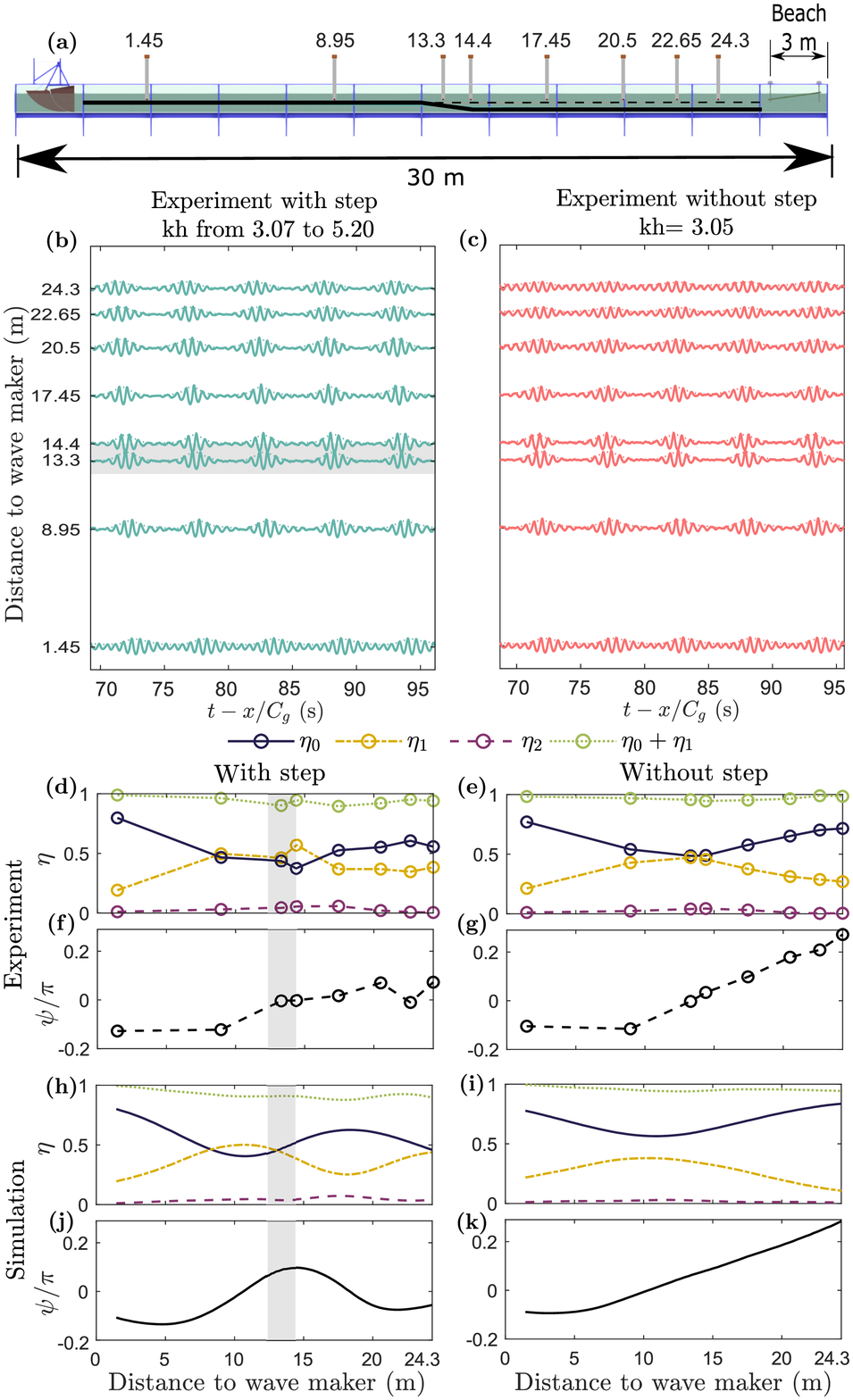}
\caption{(a) Water wave flume with artificial floor setup. One end shows the piston-type wave maker and the other end an inclined wave absorber with an artificial grass layer. Top: positions of the wave gauges.
(b) Wave height at each recorded position for the experiment with variable bathymetry, multiplied by a factor 20, the grey stripe indicates the position of the step. (c)  Wave height at each recorded position for the experiment with constant bathymetry, multiplied by a factor 20. 
(d--k) Sideband evolution of the AB-type surface water wave over the adopted bathymetry with the depth step (d,f,h,j), and the constant
flat bottom $h^0$ (e,g,i,k).  (d--g) Sideband dynamics as identified from the eight gauge measurements, connected by a linear interpolation; (h--k) corresponding NLSE-simulated evolution.  (d,e,h,i) Sideband fractions $\eta_0, \eta_1, \eta_2$ of modes at frequencies 0 (carrier), $\pm \Omega$, and $\pm 2\Omega$, respectively;
(f,g,j,k) phase $\psi$ of first-order sidebands (modes at $\pm \Omega$) relative to the carrier frequency, \textit{i.e.}, $\eta_0 \equiv |\hat{V}(\xi,0)|^2/N$, $\eta_1 \equiv (|\hat{V}(\xi,\Omega)|^2+|\hat{V}(\xi,-\Omega)|^2)/N$, and $\eta_2 \equiv (|\hat{V}(\xi,2\Omega)|^2+|\hat{V}(\xi,-2\Omega)|^2)/N$, and $\psi\equiv \frac{\phi_1+\phi_{-1}}{2}-\phi_0$, with 
$\phi_n\equiv \mathrm{Arg}[\hat{V}(\xi,n \Omega)]$, where $\hat{V}$ denotes the Fourier transform of $V$.
}
\label{fig:new_fig_sidebands}
\end{figure}

Our theoretical results allow to design an experimental realization in a the $30\times 1$~m$^2$  water wave flume of The University of Sydney [Fig.~\ref{fig:new_fig_sidebands}(a)]. Rigid aluminium plates, 2 m long each, have been lifted from the bottom of the tank to allow a flat floor with constant depth $h^0=32.4$ cm up to the distance $x = 12.35$ m and $h^\infty = 55.2$ cm from $x=14.28$ m with a constant slope inbetween. 


%

The initial conditions feature a carrier at a central frequency
$f_0=1.53$~Hz slowly modulated with frequency (sideband detuning) $f_m= 0.18$~Hz to form an AB focusing at $x = 10.28$~m \cite{chabchoub2017experiments}. These carrier and modulation frequencies are within reach of the wave maker ($f \le 2$~Hz). This implies  $\kappa^0=3.06$ and  $\kappa^\infty=5.02$, and the initial steepness is $\Eps=0.14$, most likely preventing wave breaking.
With these parameters, we obtain $\Omega^0=1.67<\sqrt{3}$, allowing us to observe one FPUT cycle within the tank length. 
Eight resistive wave gauges 
characterized the wave train evolution before, during, and after the depth transition. 
The first gauge is used to reconstruct, by conventional Hilbert-transform and bound mode filtering \cite{Osborne2009}, the envelope used for numerical integration of the NLSE [Eq.~\eqref{eq:DReq2}], including linear dissipation resulting from inclined beds \cite{HUNT1952a}.

\aasays{We compare the experimental traces with 
and without the bathymetry step. We observe that the former [Fig.~\ref{fig:new_fig_sidebands}(b)] still exhibits a train of clearly modulated pulses at the end of the tank, while the latter [Fig.~\ref{fig:new_fig_sidebands}(c)] qualitatively recurs to the initial state. This is particularly evident by comparing traces at $x = 22.65$ and $x=24.3$. This is the first strong evidence of  stabilization. } 

In order to quantitatively reconstruct the  phase-space trajectories described above and map them [Fig.~\ref{fig:principle}(d)], we directly Fourier-transform the surface elevation to extract the amplitude of the central mode and of the (unstable) $\pm \Omega$ and (stable) $\pm 2 \Omega$ sidebands [Figure~\ref{fig:new_fig_sidebands}(d)], as well as the relative phase $\psi$ between the carrier and the unstable sidebands 
[Figure~\ref{fig:new_fig_sidebands}(f)]. 
The $\pm \Omega$ sidebands grow until $x \approx 14$~m, \textit{i.e.}, where the depth step (gray band) stabilizes them to a relatively constant value, preventing the FPUT recurrence. The central mode evolves complementarily. Simultaneously, the relative phase of the first sideband pair stops growing. NLSE simulations reproduce quantitatively this behavior, with a stabilization of the sidebands to a high value and a stop to the growth of the sideband phase [Figure~\ref{fig:new_fig_sidebands}(h, j)]. This behavior contrasts with both the measurements [Figure~\ref{fig:new_fig_sidebands}(e, g)] and the simulation [Figure~\ref{fig:new_fig_sidebands}(i, k)] on a uniform depth, for which the FPUT recurrence is expected to occur before the end of the flume while the relative phase $\psi$ grows steadily. The small discrepancy between the focal point of the AB chosen as initial condition and the actual measured value ascribe to dissipation \cite{Kimmoun2016} and to higher-order physical effects, disregarded in the NLSE \cite{Zhang2014, Armaroli2017}.
\aasays{We interpret the small decay (resp. growth) of $\eta_0$ (resp. $\eta_1$) just after the depth jump as due to a partial reflection of the wave on the transition region, yielding imperfect energy transfer or to a small inaccuracy in gauge calibration.}

\begin{figure}[tb]
\centering
\includegraphics[width=\columnwidth]{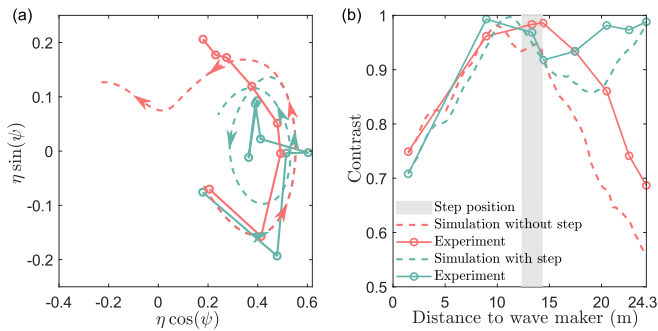}
\caption{(a) Propagation of a surface water wave AB over the depth step followed by flat bottom, displayed in the phase plane of Fig.~\ref{fig:principle}(d), where $\eta\equiv\eta_1$ of Fig.~\ref{fig:new_fig_sidebands}. The respective NLSE simulations last up to $37.7$~m. (b) Corresponding envelope contrast $C\equiv 1-\frac{\min|U|}{\max|U|}$. 
}
\label{fig:visibility}
\end{figure}

Sidebands at $\pm 2 \Omega$ stay below 6\%. Therefore, we can safely rely on the reduced set of variables introduced originally in \cite{Trillo1991c} and recently employed in nonlinear fiber optical experiments \cite{Mussot2018,Vanderhaegen20} (Supplemental Material S2~\cite{SM2020}). In Figure~\ref{fig:visibility}(a) we map the experimental trajectories onto the plane of Fig.~\ref{fig:principle}(d) and compare them to simulated results. While over a flat bottom the system is ejected outside of the separatrix and displays unlocked  phase growth, the bathymetry step forces the trajectory inside the separatrix, clearly shown by phase locking at $\psi\approx0$.

By estimating $T_\mathrm{nl}$ from the depth, carrier frequency, and the experimental value of $V_0\sqrt{N}$, we derive the normalized detuning values:  before the step, at $x=8.95$ m, $\Omega^0\approx 1.67$, while after, at $x=14.40$~m, the value of $\Omega^\infty\approx 1.34$ is indeed very close to the theoretical optimal as in Eq.~\eqref{eq:freqmatch}.

The effect of the depth step is even more visible by looking at the contrast $C\equiv 1-\frac{\min|U|}{\max|U|}$ of the temporal envelope modulation, averaged over all the modulation cycles comprised in the measured waveform [Figure~\ref{fig:visibility}(b)]. The contrast rises to 1 in the AB focusing region (``inspiration'' of the AB). On a flat bottom, it symmetrically decays after the focus (AB ``expiration'') due to the FPUT recurrence. Conversely, the bathymetry step locks the contrast to its maximum value. NLSE simulations  reproduce well this behavior. Analogous experimental results can be achieved for near-AB conditions, (Supplemental Material S4~\cite{SM2020}).

To summarize, we have found a theoretical condition to dynamically stabilize unstable nonlinear waves. While the approach applies to any system described by the NLSE, and could therefore be easily generalized to other dynamical models, we have experimentally confirmed our finding for the specific case of wave hydrodynamics. A sharp change in water depth simultaneously modifies the dispersion and nonlinearity experienced by surface gravity wave packets, thus dramatically modifying their dynamical behaviour. 
In the case of ABs, the separatrix expands and ends up enclosing the system trajectory, which is stabilized around an elliptic fixed point, \textit{i.e.}, a center.
This jump can be described as the optimal matching of an initial AB solution to a steady dnoidal solution of the universal NLSE, illustrating the generality of this wave control process. This approach contrasts with that of a slow evolution of the system over several envelope oscillations, that also results in system stabilization 
\cite{Armaroli2020}, and from stabilization mechanisms relying on dissipation~\cite{SotoCrespo2017}.

We anticipate that this cross-disciplinary approach will be further explored in other nonlinear dispersive media and will improve understanding of nonlinear wave control and transformation through a change of the waveguiding and consequently wave propagation characteristic parameters.

\begin{acknowledgments}
We acknowledge financial support from the Swiss National
Science Foundation (Project No.~200020-175697) and the University of Sydney--University of Geneva Partnership collaboration award. We thank Debbie Eeltink for fruitful discussion. 
Zachary Benitez and Theo Gresley-Daines are acknowledged for the meticulous design of the experimental set-up and technical support. \end{acknowledgments}

\bibliography{./Autoresonance,./MIandAB,./ST,./OpticalAnalogies,./VariableDepth,./HydroOther}

\end{document}


\title{
Stabilization of unsteady nonlinear waves by phase space manipulation
\\Supplementary material } 

\author{{Alexis Gomel} }
\affiliation{GAP, Universit{\'e} de Gen{\`e}ve, Chemin de Pinchat 22, 1227 Carouge, Switzerland}
\affiliation{Institute for Environmental Sciences, Universit{\'e} de Gen{\`e}ve, Boulevard Carl-Vogt 66, 1205 Gen{\`e}ve, Switzerland}

\author{{Amin Chabchoub}}
\affiliation{Centre for Wind, Waves and Water, School of Civil Engineering, The University of Sydney, NSW 2006, Australia}
\affiliation{Disaster Prevention Research Institute, Kyoto University, Kyoto 611-0011, Japan}

\author{{Maura Brunetti}}
\affiliation{GAP, Universit{\'e} de Gen{\`e}ve, Chemin de Pinchat 22, 1227 Carouge, Switzerland}
\affiliation{Institute for Environmental Sciences, Universit{\'e} de Gen{\`e}ve, Boulevard Carl-Vogt 66, 1205 Gen{\`e}ve, Switzerland}

\author{{Stefano Trillo}}
\affiliation{Department of Engineering, University of Ferrara, via Saragat 1, 44122, Ferrara, Italy}

\author{{J{\'e}r{\^o}me Kasparian}}
\email{jerome.kasparian@unige.ch}
\affiliation{GAP, Universit{\'e} de Gen{\`e}ve, Chemin de Pinchat 22, 1227 Carouge, Switzerland}
\affiliation{Institute for Environmental Sciences, Universit{\'e} de Gen{\`e}ve, Boulevard Carl-Vogt 66, 1205 Gen{\`e}ve, Switzerland}
\date{\today}

\author{{Andrea Armaroli}}
\email{andrea.armaroli@univ-lille.fr}
\affiliation{GAP, Universit{\'e} de Gen{\`e}ve, Chemin de Pinchat 22, 1227 Carouge, Switzerland}
\affiliation{Institute for Environmental Sciences, Universit{\'e} de Gen{\`e}ve, Boulevard Carl-Vogt 66, 1205 Gen{\`e}ve, Switzerland}
\affiliation{Present address: PhLAM - Laboratoire de Physique des Lasers, Atomes et Mol\'ecules, IRCICA, 50 avenue Halley, 59658 Villeneuve d'Ascq, France. }

\maketitle

\section{Derivation of Eq.~(1) and properties of its coefficients}

In this Supplemental Material, we detail the derivation of Eq.~(1) of the main text and illustrate a few properties of its coefficients.
Our specific goals are: (i) to show that the NLSE with $\xi$-dependent coefficients, denoted as Eq.~(1) in the main text, can be derived from a NLSE including an explicit shoaling term by an elementary transformation; (ii) to recall the generic trend of depth-dependent parameters.

Although this material is easily found in the literature, it is helpful to the interested reader who would like to easily find the admissible range for  the physical quantities under study. 

Applying the method of multiple scales up to $\mathcal O( \Eps^3)$ to the inviscid irrotational water wave problem yields the evolution equation \cite{Djordjevic1978,Mei2005}
\begin{equation}
i\pder{U}{\xi} + \beta\pdern{U}{\tau}{2} - \gamma|U|^2U = -i \mu U,
\label{eq:DReq1}
\end{equation}
where $U(\xi,\tau)$ is the envelope of the free surface elevation. $\xi\equiv\Eps^2 x$, and $\tau\equiv \Eps\left[\int_0^x \frac{\ud \zeta}{c_\mathrm{g}(\zeta)}-t\right]$ ($t$ being the physical time) are the coordinates in a frame moving at the group velocity of the envelope, $c_\mathrm{g} \equiv \pder{\omega}{k}= \frac{g}{2\omega}\left[\sigma + \kappa(1-\sigma^2)\right]$. $\beta$, $\gamma$, and $\mu\equiv\mu_0 \frac{\ud \kappa}{\ud \xi}$ represent the dispersion, cubic nonlinearity and shoaling coefficient, respectively.  The first two are the coefficients of the NLSE on uniform depth~\cite{Hasimoto1972}, and depend on $\kappa \equiv kh$ only~\cite{Mei2005}. Finally, $\mu$ results from wave energy conservation arguments as  $\mu_0\equiv\frac{1}{2{\omega c_g}}\frac{\ud \left[\omega c_g\right]}{\ud \kappa}$. 

\begin{figure}[tb]
\centering
\includegraphics[width=0.8\textwidth,trim= 50 180 20 10 , clip=true]{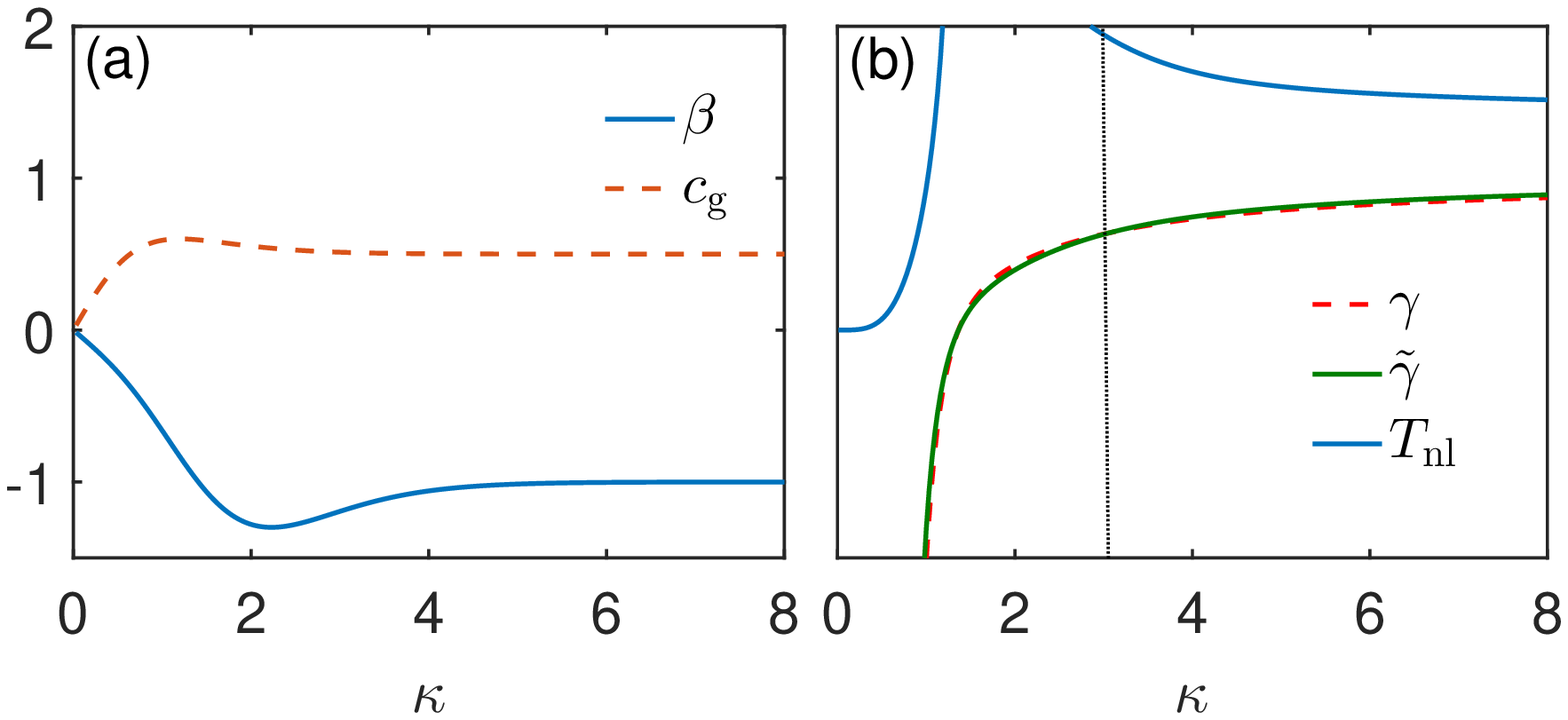}
\caption{Dependence of the coefficients of Eq.~\eqref{eq:DReq1} on the depth parameter $\kappa$, with $g=\omega=1$. (a) Dispersion parameters ($c_\mathrm{g}$: group velocity; $\beta$: group velocity dispersion); (b) Nonlinear parameters ($\gamma$: nonlinear coefficient; $\tilde\gamma$: shoaling-corrected nonlinear coefficient; $T_\mathrm{nl}$: nonlinear time, calculated for $V_0=1$). Black dotted vertical line: reference value  $\kappa(\xi=0)=3.06$ corresponding to the initial value used in the experiments. Notice the minor impact of shoaling. }
\label{fig:DRparams}
\end{figure}

Here $\beta<0$ regardless of $\kappa$, provided only surface gravity waves are considered \cite{Mei2005} [see Fig.~\ref{fig:DRparams}(a)]. $\gamma\ge0$ for $\kappa\ge1.363$ [see Fig.~\ref{fig:DRparams}(b)]. $c_\mathrm{g}$ [see Fig.~\ref{fig:DRparams}(a)] is maximum for $\kappa \approx 1.20$ (note that, for definiteness in Fig.~\ref{fig:DRparams}(a,b) we have set $g=\omega=1$).

Let  $V\equiv U\exp\left[\int_0^\xi{\mu (y) \,\ud y}\right]$, a shoaling-corrected complex amplitude~\cite{Onorato2011}. Eq.~\eqref{eq:DReq1} becomes
\begin{equation}
i\pder{V}{\xi} + \beta\pdern{V}{\tau}{2} - \tilde\gamma|V|^2V = 0,
\label{eq:DReq2}
\end{equation}
 \textit{i.e.}, a varying parameters NLSE, with $\tilde\gamma(\xi) \equiv \gamma(\xi) \frac{c_\mathrm{g}(\xi=0)}{c_\mathrm{g}(\xi)}$. 
As  $c_\mathrm{g}$ monotonically decreases in the focusing regime, $\beta\tilde\gamma<0$, shoaling slightly increases the effective nonlinearity $\tilde\gamma$ [Fig.~\ref{fig:DRparams}(b)] for $\kappa>\kappa^0$, \textit{i.e.}, if the depth increases along the propagation. 

In the main text, we derived a normalized NLSE by defining a nonlinear length $L_\mathrm{nl}$ and a nonlinear time $T_\mathrm{nl}=\sqrt{\frac{2|\beta|}{|\tilde\gamma| V_0^2}}$. The latter physically summarizes the design criterion for the bathymetry step, Eq.~(7) in the main text. The  modulation frequency $f_\mathrm{m}$ of the initial sideband perturbation is obviously constant, while the normalized angular frequency $\Omega=2\pi f_\mathrm{m} T_\mathrm{nl}$, which controls the phase-space properties (see the main text and below in \ref{sec:3wave}), has to decrease according to Eq.~(7) in the main text. In Fig.~\ref{fig:DRparams}(b) $T_\mathrm{nl}$ is shown (for $V_0=1$), that exhibits a singularity at $\kappa=1.363$ and converges to a constant value $\sqrt{2}/V_0$ for $\kappa\to\infty$.

\section{Three-wave truncation and Hamiltonian formalism}
\label{sec:3wave}

In this Supplemental Material, we detail how a three-wave truncation \cite{Cappellini1991,Trillo1991c,Armaroli2017} can also be successfully applied to study the separatrix dilation process.

This approach provides good qualitative estimates for generic initial conditions where only one unstable sideband is included in the initial conditions. Compared to exact NLSE solutions we rely on in the main text, it allows us (i) to obtain a closed form condition for optimal freezing, which, however, differs from Eq.~(8) of the main text, thus requiring to be comparatively tested;
(ii) to establish a set of reduced variables and the associated phase-plane, on which the infinite-dimensional dynamics can be projected to obtain a clear insight. This projection has been proficiently used in the main text even beyond the truncated case for the matching of exact solutions (AB, dn-oidal).

\subsection{The three-wave system for the depth-varying NLSE}
Let us substitute the Ansatz $V(\xi,\tau)= A_0(\xi) + A_1(\xi) e^{i \Omega \tau} + A_{-1}(\xi)e^{-i\Omega \tau}$ in the NLSE [Eq.~(1) of the main text]. By retaining only the terms oscillating at the frequencies $0$ and $\pm\Omega$, we obtain the three-wave system of ODEs (prime stands for $d/d\xi$)
\begin{equation}
\begin{aligned}
	iA_0' =&\tilde\gamma  (|A_0|^2 + 2 |A_1|^2 + 2 |A_{-1}|^2 )A_0 +  2\tilde\gamma   A_{1}A_{-1}A_0^*,\\
	iA_1' =& \beta\Omega^2 A_1 +\tilde\gamma  (|A_1|^2 + 2 |A_0|^2 + 2 |A_{-1}|^2 )A_1 + \tilde\gamma   A_{-1}^*A_0^2,\\
	iA_{-1}' =& \beta\Omega^2 A_{-1} +\tilde\gamma  (|A_{-1}|^2 + 2 |A_0|^2 + 2 |A_{1}|^2 )A_{-1} +  \tilde\gamma   A_{1}^*A_0^2.
\end{aligned}
\label{eq:3complexwaves}
\end{equation}

By introducing squared amplitudes $\zeta_n$ and phases $\phi_n$ of the three modes, and substituting  $A_n = \sqrt{\zeta_n}\exp i \phi_n$, $n=-1,0,1$, in Eq.~\eqref{eq:3complexwaves}, we notice that the phases appear only through the overall combination $\psi=\frac{\phi_1+\phi_{-1}}{2}-\phi_0$ defined in the main text. Moreover, it is easy to observe that the total energy $E \equiv \zeta_0 + \zeta_{1} + \zeta_{-1}$ as well  as the sideband imbalance $\chi\equiv\zeta_1-\zeta_{-1}$ are conserved. It is thus convenient to define the sideband fraction $\eta\equiv \frac{\zeta_1+\zeta_{-1}}{E}$ so that $\eta\in[0,1]$. These definitions are used for Figs.~1(d), 3 and 4 of the main text. 

Then some trivial algebra allows one to reduce Eq.~\eqref{eq:3complexwaves} as
\begin{equation}
\begin{aligned}
	\psi' &=  -\beta \Omega^2 + \tilde\gamma E \left[\frac{3\eta}{2}-1\right]+\tilde\gamma E S\cos 2\psi \left[1+\frac{\eta(\eta-1)}{S^2}\right]\\
	\eta' &= 	2\tilde\gamma E S(\eta-1)\sin2\psi,
\end{aligned}
\label{eq:3wavesetapsi}
\end{equation}
with $S = \left[(\eta-\tilde\chi)(\eta+\tilde\chi)\right]^{\frac{1}{2}}$, where $\tilde\chi \equiv \frac{\chi}{E}$ is the normalized imbalance.\\
The system~\eqref{eq:3wavesetapsi}  is integrable in Liouville sense.
For the sake of simplicity, we consider only the case of balanced sidebands $\tilde\chi=0$, yielding $S=\eta$. This is the most relevant for our theory and experiments and allows us to cast Eqs.~\eqref{eq:3wavesetapsi} in the following Hamiltonian form
\begin{equation} \label{eq:3wave1}
\psi' = \pder{H^{(\xi)}}{\eta};\;\eta' = -\pder{H^{(\xi)}}{\psi};\;\;H^{(\xi)}(\psi,\eta) \equiv \tilde \gamma E \eta(\eta-1)\cos 2 \psi + \tilde \gamma E \left(\frac{3\eta^2}{4}-\eta\right) - \beta\Omega^2 \eta
\end{equation}

Further, by letting $X \equiv E \int_{0}^\xi{ \tilde \gamma (y)\ud y}$, the system~\eqref{eq:3wave1}
can be recast in the following final form of 1 d.o.f. integrable system,  
\begin{equation}
\dot\psi = \pder{H^{(X)}}{\eta};\;\dot\eta = -\pder{H^{(X)}}{\psi}\;;H^{(X)}(\psi,\eta \vert \alpha)= \eta(\eta-1) \cos 2\psi + \alpha \eta + \frac{3}{4}\eta^2,
\label{eq:3wave2}
\end{equation}
where the dot denotes the derivative with respect to $X$. 

The dynamics of Eq.~\eqref{eq:3wave2} in the phase-plane is governed by the single parameter $\alpha$ which accounts for the normalized frequency detuning of the sidebands
\begin{equation}
\alpha\equiv-\left[\frac{ \beta \Omega^2}{\tilde\gamma E}+1\right]=\left(\frac{\Omega}{\Omega_\mathrm {M}}\right)^2-1=-4 a_\mathrm{AB}+1,
\end{equation}
where $\Omega_\mathrm{M}\equiv \sqrt{\left|\frac{\tilde\gamma}{\beta}\right|}V_0$ is the dimensional detuning of the peak MI gain and $a_\mathrm{AB}$ the well known parameter of the AB.
In terms of this parameter, modulational instability (MI) occurs for $|\alpha| \le 1$, with the peak gain being obtained for $\alpha=0$, whereas $\alpha=1$ corresponds to the MI frequency cut-off.

\subsection{Fixed points}

The bifurcation analysis of Eq.~\eqref{eq:3wave2} shows the existence of equilibria or fixed points ($\eta_e, \psi_e$): 

\begin{enumerate}
\item  $\eta_e=0, \;\psi_e=\frac{\cos^{-1} \alpha}{2}$ (\textit{i.e.}, the background, a saddle point in the MI regime $|\alpha|<1$, a center otherwise);

\item  $\eta_e=\eta_e^{\mathrm{lock}}=\frac{2(1-\alpha)}{7},\; \psi_e=0,\pi$ (\textit{i.e.}, phase locked eigenmodes, centers (elliptic points) for $|\alpha|\le 1$); 

\item  $\eta_e = 1, \; \psi_e=\frac{\cos^{-1} \left(-\alpha-\frac{3}{	2}\right)}{2}$ (\textit{i.e.}, sideband mode, a center for $\alpha>-\frac{1}{2}$, a saddle otherwise);

\item $\eta_e=2(1+\alpha), \; \psi_e = \frac{ \pi}{2}+m\pi$, (\textit{i.e.}, centers for $-1<\alpha\le-\frac{1}{2}$). 
\end{enumerate}

The fixed points directly relevant to the mechanism of separatrix dilation are the first two. The first one represents the unstable background which, in phase-plane, is the saddle (in the MI-unstable range of frequencies), from which a separatrix emanates. The separatrix is nothing but the three-mode approximation of the AB.  The second one stands for the three-mode truncation of two, mutually out of phase, dn-oidal waves into which the dynamics could be stabilized. 
The third and fourth fixed points are not relevant for the present dynamics. Indeed, the sideband eigenmode is relevant only for the opposite situation of a small central frequency amplified at the expense of strong sidebands, not considered here. Furthermore, the last fixed point exists only in the lower half of the MI gain curve ($\Omega\le\frac{\Omega_\mathrm{C}}{2}$), where the three-wave truncation obviously breaks down since higher-order MI set in, which is ruled by additional unstable sideband pairs at $\pm n\Omega$, $n\in \mathbb{Z}$.

\begin{figure}[th]
\centering
\includegraphics[width=0.6\textwidth]{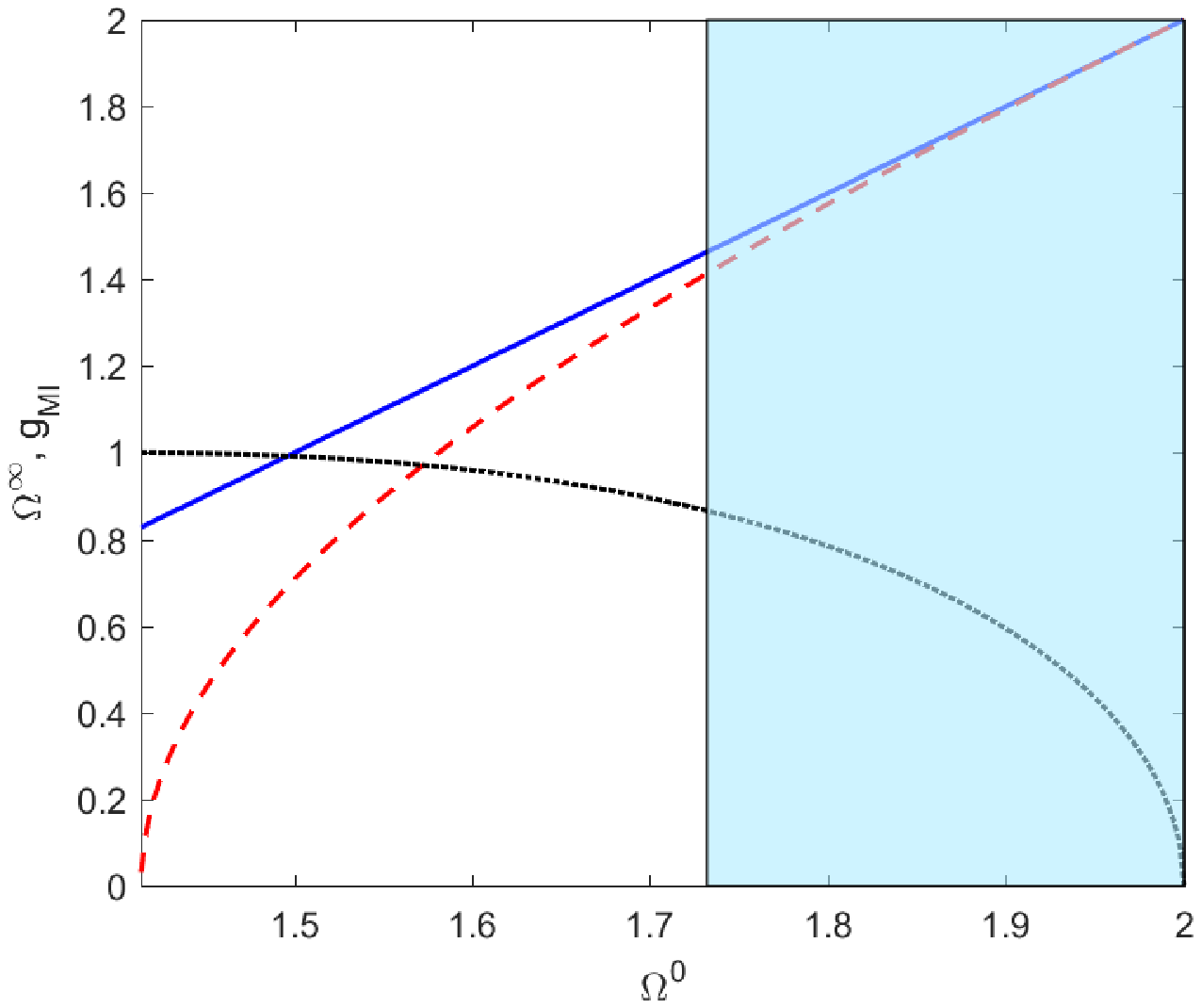}
\caption{Comparison of the matching condition based on the full NLSE solutions (AB, dn-oidal; Eq.~(8) in the main text, solid blue line) with the matching condition in Eq.~\eqref{eq:alphamatch} obtained in the three-wave approximation (red dashed line). The dotted black line corresponds to the MI gain and the light blue shaded region corresponds to the accessible region for stabilization determined in the main text, \textit{i.e.}, $[\sqrt{3},2]$. 
}
\label{fig:ABdnvs3waves}
\end{figure}

\subsection{Three-wave stabilization condition}

The condition for optimal stabilization based on the three-wave truncation can be obtained by means of a simple argument, as follows. We assume to start on the separatrix characterized by level $H^{(X)}(\eta,\psi \vert \alpha^0)=0$ with $\alpha = \alpha^0$ (here, subscripts $0,\,\infty$ denote the value of $\alpha$ before and after the jump, as in the main text). 
We impose that the change of depth occurs at the farthest point along the orbit (corresponding to maximum amplification and  compression in the AB), which can be easily found to be  $\eta=\eta_{max}=\frac{4(1-\alpha^0)}{7}$. Then, we look for the value $\alpha = \alpha_{\infty}$ for which $\eta=\eta_\mathrm{max}$ becomes an elliptic point $\eta_e^{\mathrm{lock}}$ of the new Hamiltonian $H^{(X)}(\eta,\psi \vert \alpha_{\infty})$ after the transition. We obtain the following relation

\begin{equation}
    \alpha^\infty = 2\alpha^0-1.
    \label{eq:alphamatch}
\end{equation}

Compared with Eq.~(7) of the main text, Eq. \eqref{eq:alphamatch} establishes a different relation between the normalized frequencies before and after the jump.
In Fig. \ref{fig:ABdnvs3waves} we contrast how $\Omega^\infty$ depends on $\Omega^0$ as prescribed by the two matching conditions [Eq.~(7) and Eq.~\eqref{eq:alphamatch}], in the spectral region of interest (shaded domain). As shown, Eq.~\eqref{eq:alphamatch} implies a smaller $\Omega^\infty$, thus requiring a larger depth jump.

We also emphasize that, when the jump is designed to fulfill Eq.~\eqref{eq:alphamatch}, the three-wave dynamics gives rise to an ideal freezing into the fixed point. However, even under the validity of Eq.~\eqref{eq:alphamatch}, the freezing is no longer ideal in the full NLSE dynamics, due to the impact of higher-order harmonics.
This is shown in the next section, see Fig.~\ref{fig:simulatedPP}.

\section{Additional numerical results}

In this Supplemental Material, our aim is to present more extensive numerical results to illustrate in more detail the principle of stabilization. 
In particular, our specific goals are: (i) to explore the dynamics for larger distances compared with those available in the experimental facility; 
(ii) to compare the dynamics under the different matching conditions [Eq.~(7) in the main text based on AB-dnoidal matching versus Eq.~\eqref{eq:alphamatch} based on 3-wave truncation];
(iii) to show that the principle of stabilization is not limited to initial conditions lying strictly on the separatrix or AB, but works equally well for near-separatrix dynamics, too. To this purpose, a  three-wave (carrier and a first-order sideband pair) initial condition is chosen with a relative phase adjusted to lie either outside or inside the separatrix.

\begin{figure}[t]
\centering
\includegraphics[width=0.5\textwidth]{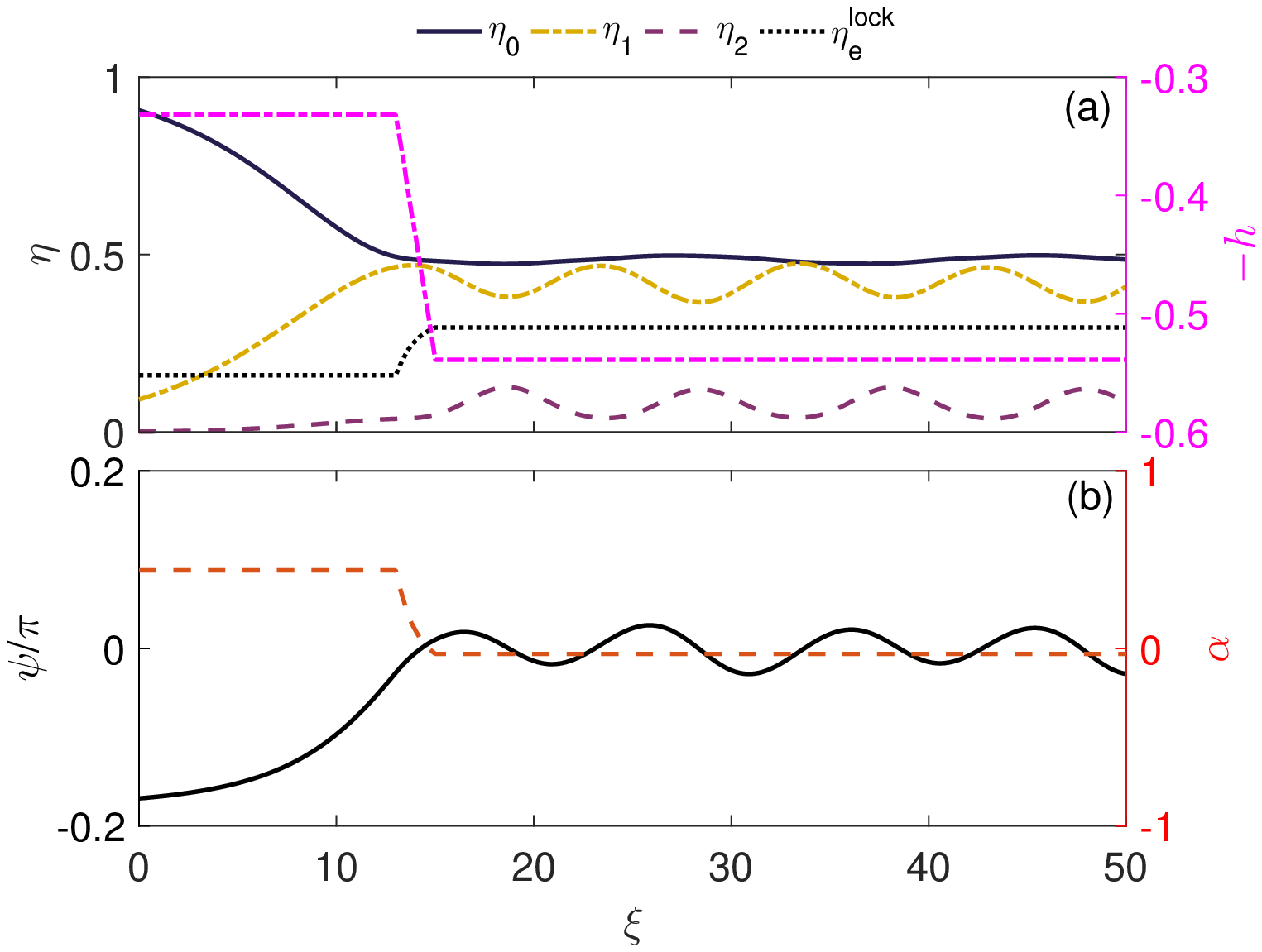}
\caption{
Results of the numerical integration of the NLSE with same parameters as in Fig. 4 of the text and longer distance. In panel (a) we show the evolution of sideband norm fractions of the background $\eta_0$, first-order sidebands $\eta_1 \equiv \eta $, and second-order sidebands $\eta_2$. The fixed point $\tilde\eta_2$ predicted by the three-wave is shown as a black dotted line. The dash-dotted purple line shows the evolution of depth (units on the right vertical axis). Panel (b) shows the relative phase [as in is the Figs.~4(c-d) of the main text], as well as of the MI parameter $\alpha$ defined for the three-wave system (red dashed line, units on the right vertical axis).
}
\label{fig:simulatedsidebandsAB}
\end{figure}

First, we simulate identical frequencies and jump condition as in the main text [$\Omega^0=1.67$, jump determined by Eq.~(7)]. 
We use as initial condition an AB focusing at $\xi=14$~m  as in the experiment; no damping is included. However the total propagation distance is chosen to be longer (up to 50 m) to appreciate the overall effect of the jump in the long range.

As shown in Fig.~\ref{fig:simulatedsidebandsAB}(a), the stabilization obtained under the optimal condition [Eq.~(7)] is fairly good. 
The background component stays nearly constant after the jump, while the first- and second-order sidebands exhibit spurious residual quasi-periodic oscillations, which turn out to be mutually out of phase. Figure~\ref{fig:simulatedsidebandsAB}(b) confirms (as already noticed in Fig.~4(f,g) of the main text that, after the jump, the phase exhibits small oscillations around the phase of the elliptic point  $\psi_e = 0$ (see solid  black line). In order to compare with the three-wave approach, we have also reported in Fig. \ref{fig:simulatedsidebandsAB}(a), as a  dotted black line, the sideband fraction of the elliptic point $\eta_e^{\mathrm{lock}}$. 
The discrepancy which can be noticed in Fig.~\ref{fig:simulatedsidebandsAB}(a) between the average value of $\eta$ obtained from numerical simulations and $\eta_e^{\mathrm{lock}}$
must be ascribed to the loss of accuracy of the three-wave truncation \cite{Trillo1991c,Armaroli2017}. 
The method is indeed less and less accurate as $\alpha$ approaches $0$. Given $\alpha^0= 0.396$, Eq.~\eqref{eq:alphamatch} yields $\alpha^\infty=-0.208$, instead of the optimal value shown in Fig.~\ref{fig:simulatedsidebandsAB}(b), \textit{i.e.}, -0.05.

\begin{figure}[tb]
\centering
\includegraphics[width=0.6\textwidth]{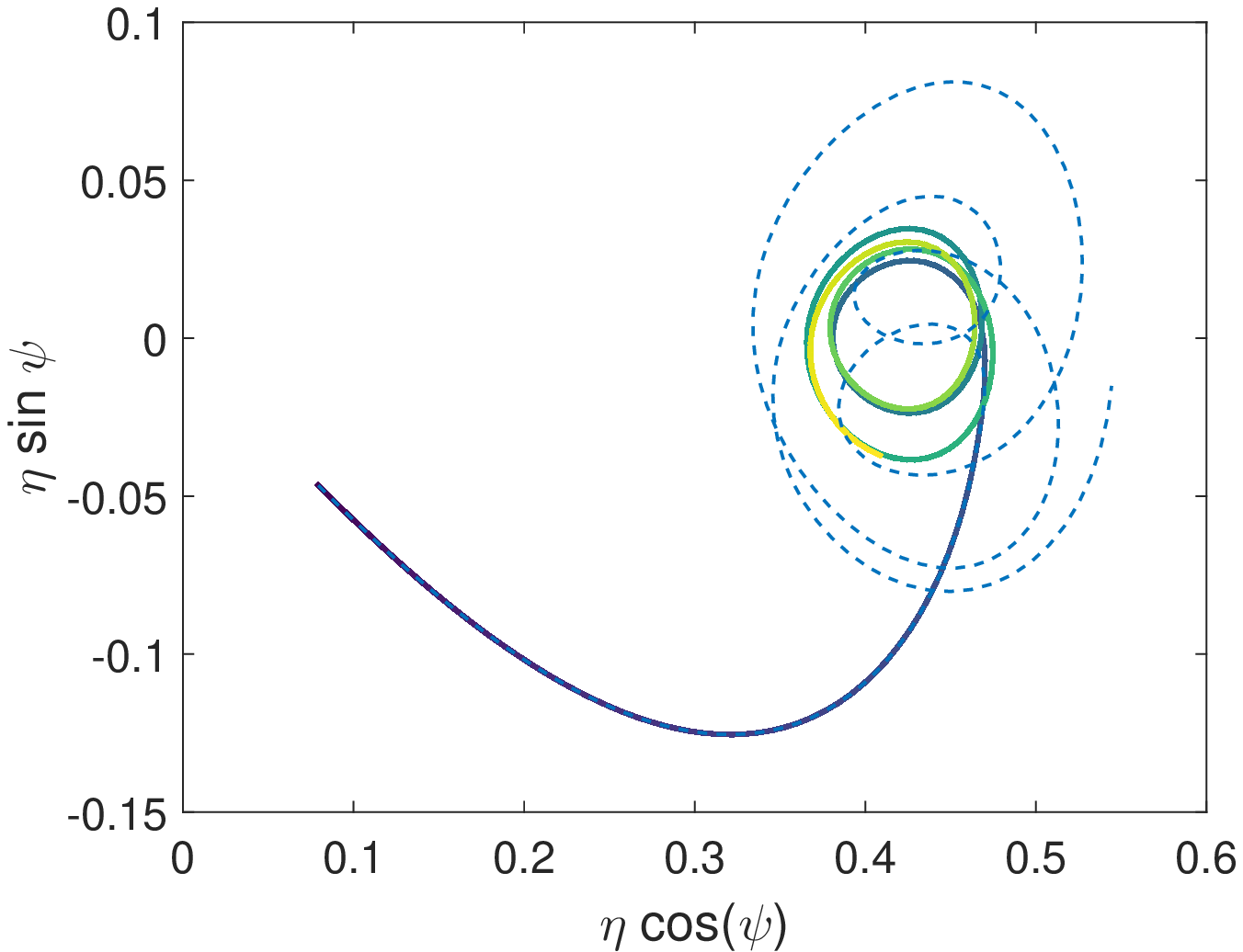}
\caption{
Results of the numerical integration of the NLSE projected on the three-wave phase plane ($\eta \cos \psi, \eta \sin \psi$), with initial condition on the AB. We compare the case ruled by the matching condition in Eq.~(7) of the main text (solid line with changing hue) with the analog simulation (dashed cyan line) for the step condition corresponding to Eq.~\eqref{eq:alphamatch},\textit{ i.e.}, from $\kappa^0=3$ to $\kappa^\infty=7$. 
}
\label{fig:simulatedPP}
\end{figure}

Second, we compare in Fig.~\ref{fig:simulatedPP} the stabilization dynamics ruled by the full NLSE for two different jumps, chosen to obey Eq. (8) of the main text or Eq.~\eqref{eq:alphamatch}, respectively. In order to have a graphic phase-plane representation, we extract from the NLSE dynamics the main Fourier components (background wave and first-order sidebands) and project the evolution on the reduced phase-plane ($\eta \cos \psi, \eta \sin \psi$). The solid line with changing hue displays the projection that arises for the matching in Eq.~(8) (same run as in  Fig.~\ref{fig:simulatedsidebandsAB}). As shown, the evolution starts near the saddle in the origin and evolves around the unstable manifold of the separatrix up to the maximum growth of the sideband where the depth jump sets in. After the transition, only quasi-periodic small oscillations around the stable equilibrium occur. Importantly, however, such oscillations become much stronger for the case in which the jump fulfils the three-wave condition Eq.~\eqref{eq:alphamatch} (see dashed line in Fig.~\ref{fig:simulatedPP}; clearly up to the jump the dynamics  is indistinguishable in the two cases). Therefore, we conclude that the recipe given by Eq.~(8) is more accurate to achieve stabilization via separatrix dilation, compared with the simplified constraint  in Eq.~\eqref{eq:alphamatch}.

\begin{figure}[tb]
\centering
\includegraphics[width=0.48\textwidth]{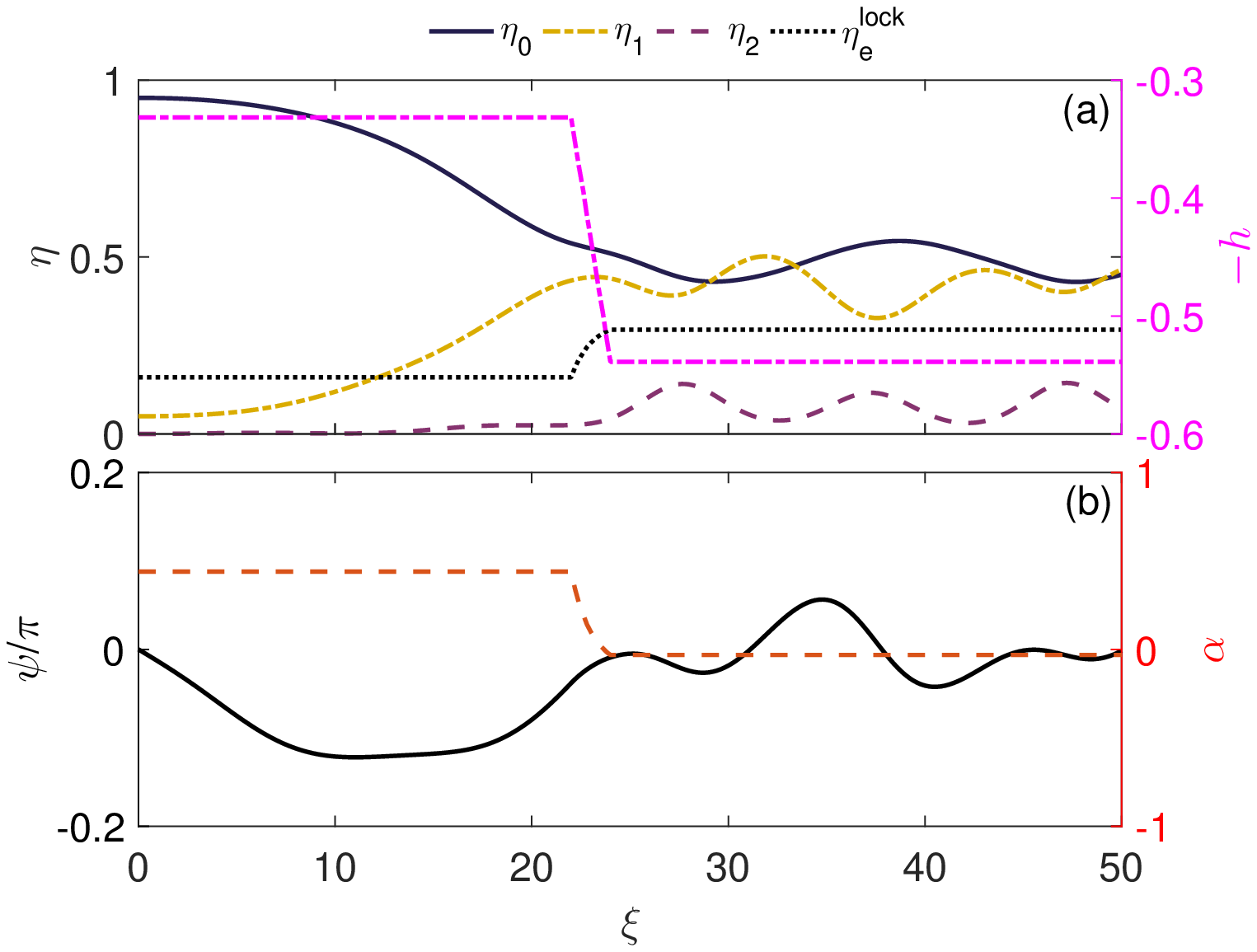}
\includegraphics[width=0.48\textwidth]{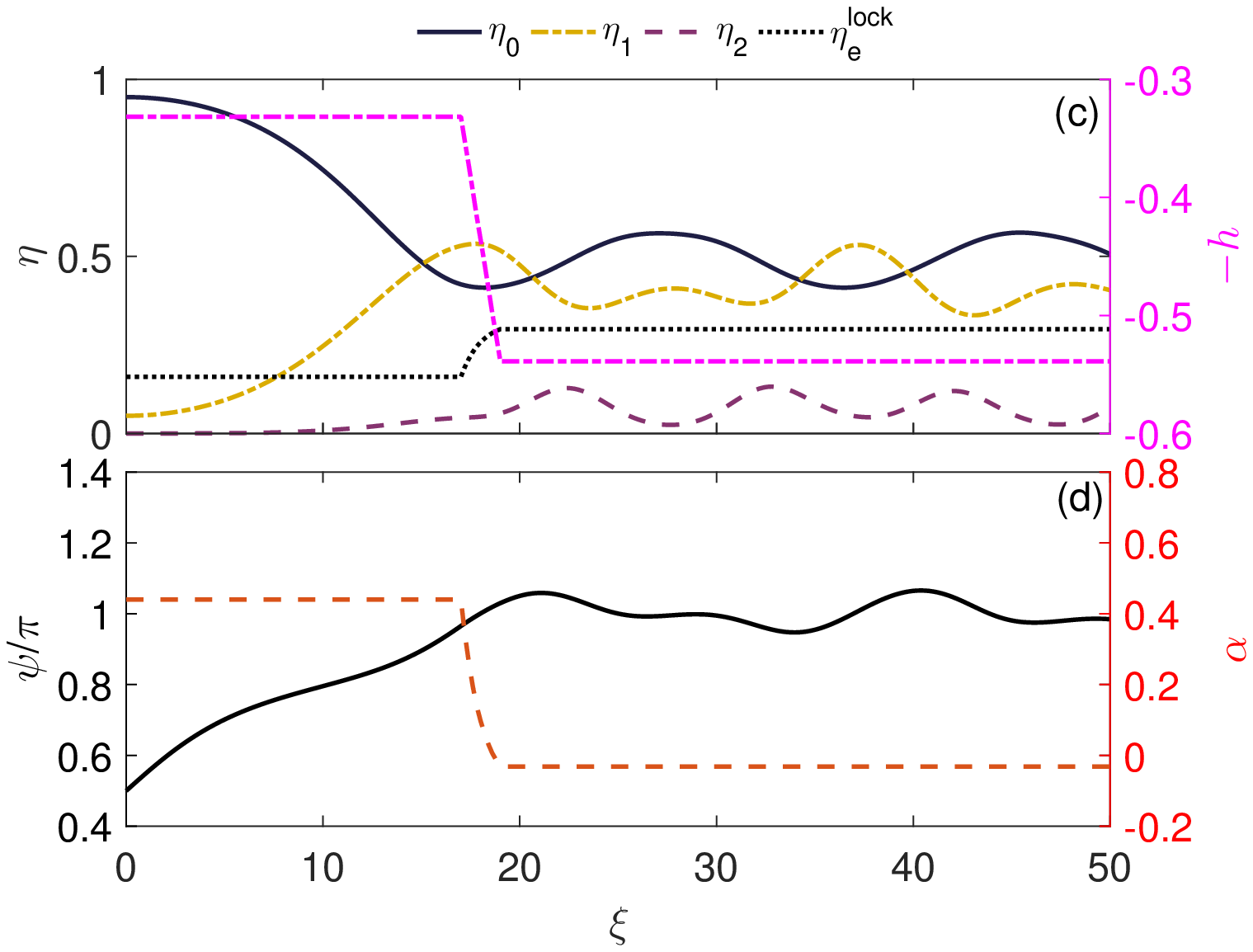}
\caption{Same as Fig.~\ref{fig:simulatedsidebandsAB} for three-wave excitation with a pair of symmetric sidebands with sideband fraction $\eta_0=0.05$.
(a,b) input amplitude modulation or in-phase case $\psi_0=0$, leading to inner near-separatrix orbit;  
(c,d) input frequency modulation, $\psi_0=\pi/2$, leading to outer near-separatrix orbit.
}
\label{fig:3wavesinside}
\end{figure}

Finally, in Fig.~\ref{fig:3wavesinside} we report evidence that the stabilization scheme based on separatrix dilation is effective also when the initial condition is not chosen strictly on the AB,
but rather involves near-separatrix orbits which give rise to periodic evolutions (Fermi-Pasta-Ulam dynamics) in the absence of the depth jump. To this end, we report examples of NLSE evolution obtained by starting with three-waves with 5\% input fraction (in the norm sense) in the sidebands and  $\Omega^0=1.67$ as above. The position of the step is adjusted to correspond to the maximum conversion distance and its  amplitude is according to Eq.~(7). We choose the input phase in order to have both inner ($\psi_0=0$, see in Fig.~\ref{fig:3wavesinside}(a-b)) or outer ($\psi_0=\pi/2$, see Fig.~\ref{fig:3wavesinside}(c-d)) near-separatrix orbit. In both cases, the simulations show that a fairly good stabilization can be achieved. The discrepancies can be ascribed to the fact that exact solutions of the NLSE predict finite sideband amplitudes at $\pm n\Omega$, while we consider only a three-wave initial condition and the perturbation is non-negligible.
In Supplemental Material S3, the experimental test performed for $\psi_0=\pi/2$ are discussed.

\section{Additional experimental details and results
}

In this Supplemental Material, we present further additional measurements performed with three-wave input, in order to provide more details on the validity of our approach under more general initial conditions.

The experimental facility is the same as in the main text, sketched in Fig.~2(a).
The amplitudes of Fourier components in the main text are obtained through Fourier analysis and post-processing of these traces.

\begin{figure}[htb]
\centering
\includegraphics[width=\columnwidth]{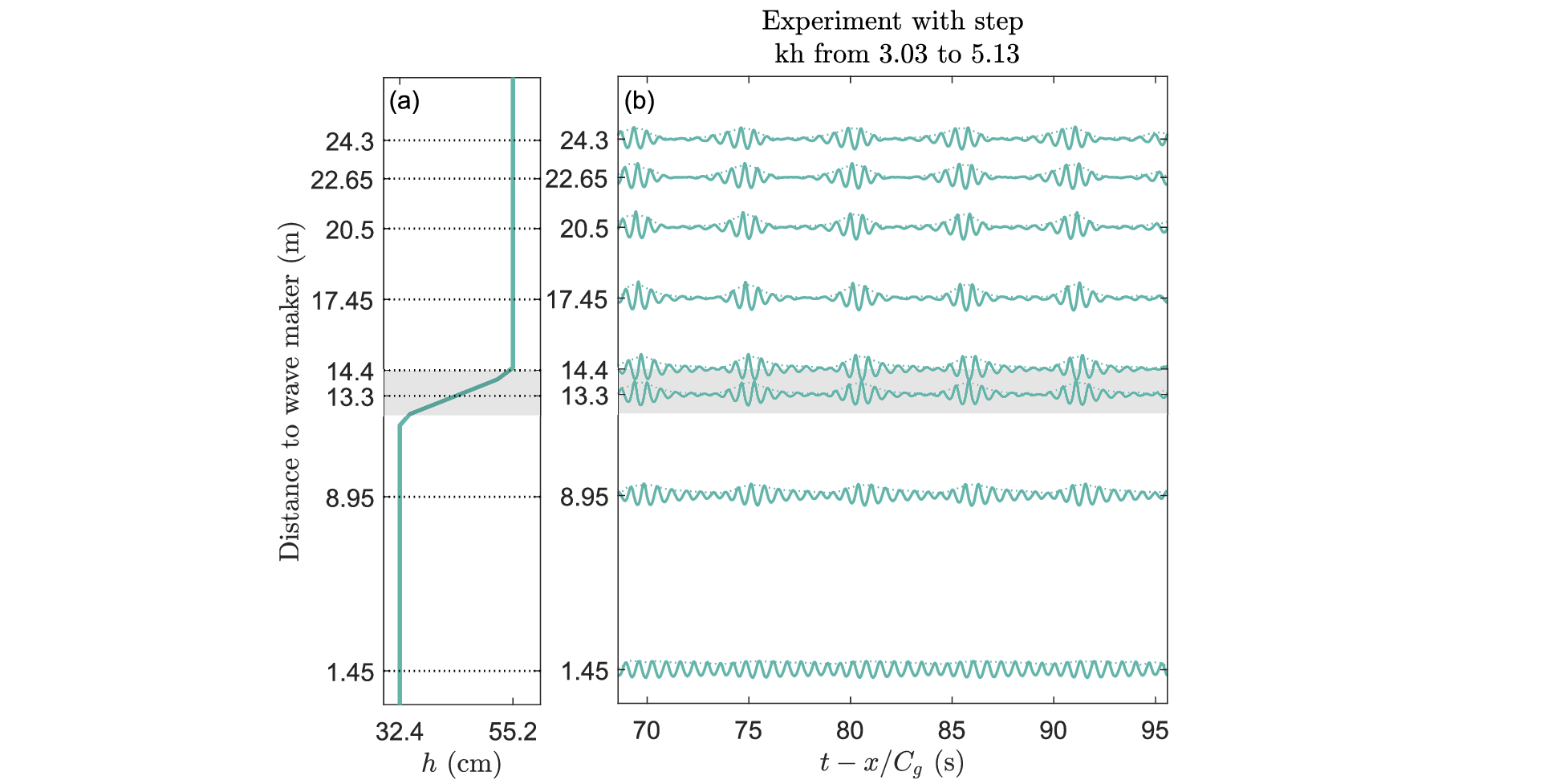}
\caption{Surface elevation and envelope for the three-wave excitation. (a) Bathymetry used in the experimental setup, the position of the wave-gauges are shown in black dotted lines. (b) Wave height at each recorded position for the experiment with a step in bathymetry, multiplied by a factor 20, the grey stripe indicates the position of the step.
}
\label{fig:cg_frame}
\end{figure}
\begin{figure}[htb]
\centering
\includegraphics[width=0.75\textwidth]{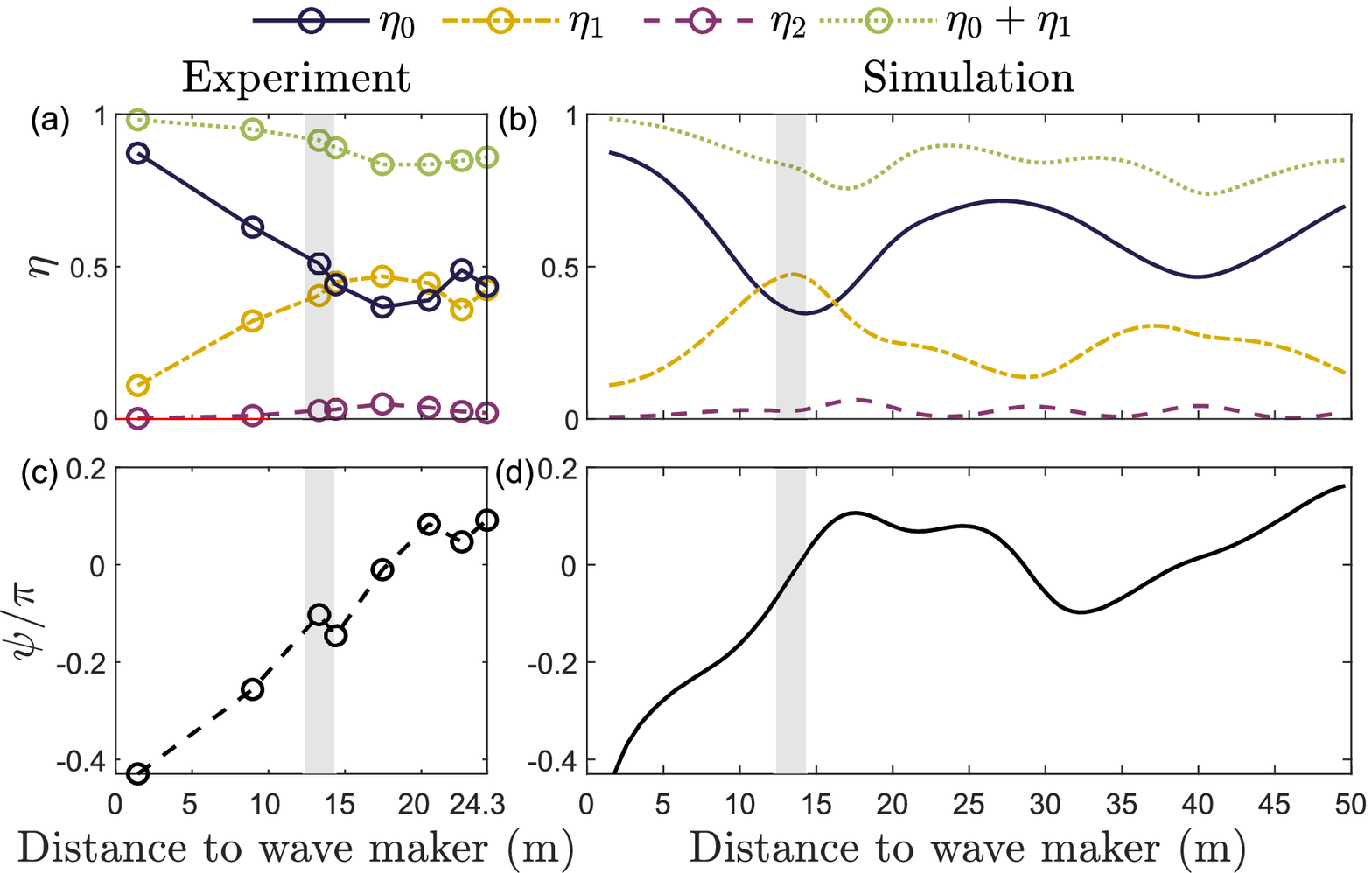}
\caption{ (a,c) Experimental and (b,d) simulated evolutions of (a,b) 
the sideband fractions $\eta_0, \eta_1, \eta_2$ of modes at frequencies 0 (carrier), $\pm \Omega$, and $\pm 2\Omega$, respectively,
and (c,d) the phase of the first sidebands with respect to the carrier.
The NLSE simulations include here damping and are extended to twice the length of the tank to illustrate the effectiveness of our approach.
}
\label{fig:sidebands3w}
\end{figure}
\begin{figure}[htb]
\centering
\includegraphics[width=0.75\textwidth]{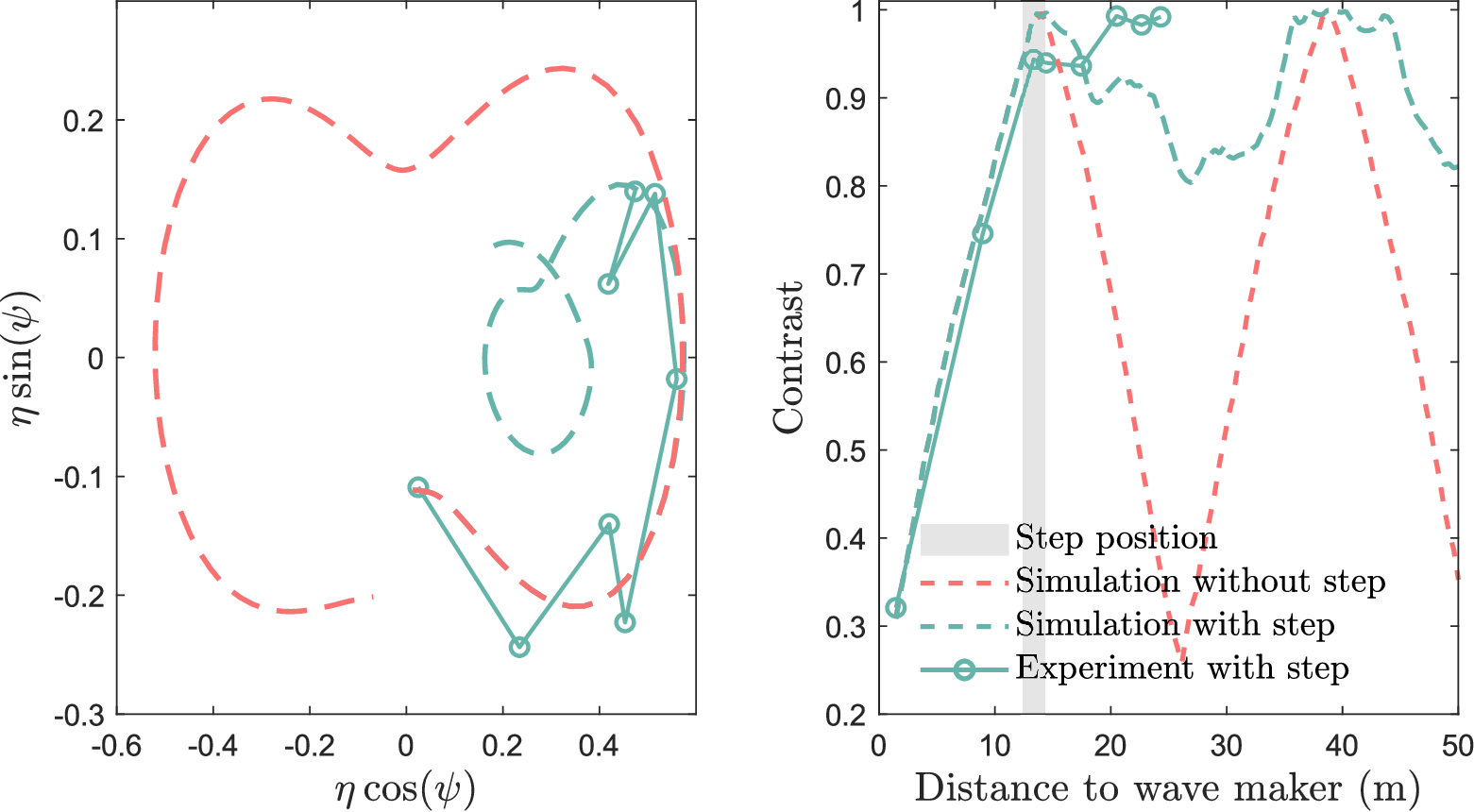}
\caption{ 
(a) Projection over the phase plane ($\eta\cos\psi,\eta\sin\psi$ of experimental trajectories in Fig.~S6-S7 (green open circles and solid line) compared with NLSE simulations with the step (green dashed line) and flat bottom (for comparison; red dashed line), respectively. The simulations last up to $50$~m. (b) Corresponding evolutions of envelope contrast as defined in the main text.
}
\label{fig:visibility3w}
\end{figure}

Then, we  report in Figs.~\ref{fig:cg_frame}, \ref{fig:sidebands3w}, \ref{fig:visibility3w} an example of experimental results achieved by exciting only three harmonic components, instead of the temporal profile of the AB.  This experiment is carried out using an initial value of $\eta_0=0.11$ and an initial relative phase of $\psi=\pi/2$, where both sidebands have the same amplitude. The carrier frequency is $f_0=1.52$~Hz and the modulation frequency (sideband detuning) is $f_m=0.18$~Hz, resulting in a nominal input dimensionless detuning $\Omega^0=1.7$ and a steepness $\epsilon=0.14$. This results in a variation of $\kappa$ from $3.06$ to $5.02$.
The simulations included for comparison a dissipation coefficient of $0.0046 \;\mathrm{m}^{-1}$ according to Ref.~\cite{HUNT1952a}, which gives a good agreement with norm damping, and results in a decay of approximately $10\%$ between the first and last gauges. This is the same value used in Figs.~2-3 in the main text. 

The experimental value of normalized frequency ranges from  
$\Omega_{0}=1.73$, as measured at the $8.95$ m gauge before the step (in order to reduce impact of damping, which reduces the effective nonlinear coefficient), and 
$\Omega_{\infty}=1.48$,
as measured at the $14.40$ m gauge, right after the step. According to the theory exposed in the main text, the expected optimal for the freezing condition is very close: $\Omega_{\infty}=1.45$.

Comparing Figs.~\ref{fig:sidebands3w} and  \ref{fig:visibility3w} to Figs.~2 and 3 in the main text, we notice that the behavior is qualitatively similar and the main conclusion is that freezing can be achieved also by deviating from the AB initial conditions. On the other hand, the discrepancy between experiments and NLSE simulations shown in Fig. \ref{fig:sidebands3w} consists mainly in a shift forward (along the tank) of the point of maximum conversion to the sidebands. This can be ascribed to higher-order effects and dissipation, which causes, in the regime of large input sidebands, deviations from the ideal phase-plane behavior. Indeed in this case the initial sideband amplitude represents a strong deviation compared with the value of the AB solution.


\bibliography{./MIandAB,./Autoresonance,./OpticalAnalogies,./VariableDepth,./HydroOther}